\documentclass{aa}  

\usepackage{txfonts}
\usepackage{graphicx}
\usepackage[utf8]{inputenc}
\usepackage{graphicx}
\usepackage{multirow}
\usepackage{amsfonts}
\usepackage{adjustbox}
\usepackage{placeins}
\newcommand{\ra}[1]{\renewcommand{\arraystretch}{#1}}
\usepackage{booktabs}
\usepackage{hyperref}
\usepackage{bm}
\usepackage{stfloats}

\def\CN2{\mbox{$C_N^2 $}}

\def\see{\mbox{$\varepsilon\ $}}

\begin{document}

   \title{Establishing baseline model performances for optical turbulence forecasting}

   \titlerunning{Establishing baseline model performances for optical turbulence forecasting}
   \authorrunning{M. De Sepibus et al.}

\author{M. de Sepibus\inst{1,2} \fnmsep\thanks{\email{marlene.desepibus@inaf.it}}
   \and 
   E. Masciadri\inst{1,2}\fnmsep\thanks{\email{elena.masciadri@inaf.it}}
   \and C. Weinberger\inst{1,2}  \and  C. Veillet\inst{3} \and S. Ragland\inst{3}}
   \institute{
   INAF Osservatorio Astrofisico di Arcetri, Largo Enrico Fermi 5, I-501 25 Florence, Italy \and
   ADONI: ADaptive Optics National laboratory of INAF \and
 Large Binocular Telescope, 933 N. Cherry Drive, Tucson, AZ, USA, 85721}

   \date{}

  \abstract
   {Accurate optical turbulence (OT) forecasts help telescopes observe with maximum efficiency, as the performance of adaptive optics (AO) systems strongly depends on turbulence conditions. Due to the chaotic and non-linear nature of turbulence, OT forecasting remains a challenging problem. A wide range of methods has been explored to address this problem, including numerical weather prediction (NWP) models, machine learning algorithms, and hybrid approaches—each varying in complexity, computational cost, and accuracy.}
   {Assessing the performance of forecasting methods is essential to benchmark them against simple, well-defined reference models, known as baselines, which establish reference thresholds. This paper aims to calculate these thresholds for the astroclimatic and atmospheric parameters most relevant for ground-based astronomy related to two of these baseline methods that are suitable for use in two different  types of forecast, which are among the most relevant for application to the ground-based astronomy: the forecast of the average of a parameter on a defined timescale and the forecast of the temporal evolution of a parameter on a defined timescale.}
   {We apply these baseline methods to a dataset related to a rich statistical sample of many years encompassing key astroclimatic and atmospheric parameters above the sites of the Very Large Telescope and the Large Binocular Telescope, and we discuss the implications of these results and the possibility of extrapolating such values for general rules. Additionally, we analyse the impact of data pre-processing steps, including smoothing, outlier rejection, and sampling frequency, on forecasting performance.}
   {We demonstrate that the predictive performance of each reference method depends on the observing sites, the parameter that is forecasted, the forecast timescale, and the type of forecast used. That means that it is meaningless to quantify the performance of OT forecasts in absolute terms, since they depend on the context.}
   {}


   \maketitle

\section{Introduction}
\label{intro}

Ground-based astronomical observations are limited by atmospheric optical turbulence, which perturbs wavefronts from observed objects and degrades the image quality at the focus of telescopes. Although adaptive optics (AO) systems are able, at present, to mitigate optical turbulence (OT) effects efficiently, their performances are strongly dependent on OT conditions.
Efficient telescope operation requires scheduling strategies that align scientific objectives with atmospheric conditions. Forecasting OT allows for proactive observation planning, enabling instruments to operate at or near peak efficiency by allocating time to programs based on expected atmospheric quality. \\
We know that optical turbulence can be forecast and different methods have been used so far for this function, including parameterization of the OT inside numerical atmospherical models  \citep{masciadri1999,cherubini2008,masciadri2017,basu2020}, empirical methods \citep{trinquet2007,giordano2013,osborn2018}, machine learning algorithms \citep{milli2020,bolbasova2021,cherubini2022,masciadri2023,li2023}, and hybrid methods that join the numerical approach with an autoregression algorithm \citep{masciadri2020,masciadri2023}. Many research groups have developed and are developing forecasting tools for OT, i.e. \CN2 profiles from which integrated astroclimatic parameters can be derived. The key question is when we can consider that a method provides good forecast performance. 

When we mention `model performances`, it means that we are interested in assessing the accuracy of the model. The accuracy of an estimate refers to how close the estimate is to the true value. However, it is not obvious how to define a 'true value', particularly for a non-linear and stochastic parameter such as optical turbulence. The most realistic way to define accuracy is to associate it with the uncertainty of measurements of the same parameter taken simultaneously with different instruments. There is an extensive discussion of this concept in \citep{masciadri2023} and we refer the reader to that paper for further information. The most common practice is to associate with the true value the values obtained by a specific instrument that is chosen as a reference. In this paper we follow this approach. It is also worth highlighting that forecast performances depend on the forecast timescale. Indeed, we can investigate the performances on short and long timescales. 

In this paper, we present two simple and well-defined forecast models that depend only on observations and can be easily implemented, which we consider as baselines for short-term models. They are computationally inexpensive methods that provide a reference level of forecast accuracy that we can consider as a threshold. If a new method (no matter the nature of the method, i.e. numerical, machine learning, or other) fails to outperform the baseline, we can conclude that the method does not provide any added value in terms of accuracy; therefore, it cannot be considered effective. These baseline methods, which we call precasting (PC hereafter) and prevision by persistence (PP hereafter), are our references in two different  types of forecast: the forecast of the average of a parameter in a specific interval of time (for the PC) and the forecast of the temporal evolution of the parameter on a specific interval of time (for the PP). In both cases, we are considering a forecast timescale of the order of one or two hours, as these are the most relevant time intervals for the science operation in ground-based astronomy. We applied the methods PC and PP above two astronomical sites, Cerro Paranal (Chile), site of the Very Large Telescope (VLT) and Mt. Graham (US), site of the Large Binocular Telescope (LBT), and we calculated the models' performances, i.e. the forecast accuracy obtained by PC and PP, for the most important astroclimatic parameters: seeing~-~$\varepsilon$, wavefront coherence time~-~$\tau_0$, isoplanatic angle~-~$\theta_0$, and ground layer fraction (GLF) and the atmospheric parameters relevant for ground-based astronomy: wind speed (WS), absolute temperature (T), relative humidity (RH), and precipitable water vapour (PWV) when observations are available in long-term datasets.

The interest in this study is driven by distinct motivations:
\begin{enumerate}
\item To estimate the baseline references, above two sites hosting the present-day top-class telescopes (VLT and LBT). Such an analysis is done with a very rich statistical sample of observations (of the order of several years) to appreciate the typical values and uncertainties in such estimates. We selected the sites of  LBT and VLT as our group developed two operational forecast systems of OT and atmospheric parameters for these two telescopes. The projects of operational OT forecasts are: (A) ALTA Center\footnote{\href{http://alta.arcetri.inaf.it}{http://alta.arcetri.inaf.it}} at LBT and (B) FATE at VLT \citep{masciadri2023}; 
\item To produce evidence, where possible, of the parameters or elements on which the performance of PC and PP depends. Possible elements that might play a role include the site, the parameter of interest (seeing, wavefront coherence time, and so on), and the definition of forecast timescale; 
\item To verify whether, by knowing the forecast performances of PC and PP methods above two sites for a specific parameter, it is possible to retrieve forecast performances of PC and PP for the same parameter above a third site. This has been done for seeing, as it is the most critical parameter for ground-based astronomy. To find an answer for this last goal, we used observations of the seeing related to a few further sites: Roque de Los Muchachos Observatory (Canary Islands), La Silla Observatory (Chile), Matera (Italy), Cagliari (Italy), and Kryoneri (Greece). The latter three sites are not astronomical sites. Matera is the location of the Space Centre Laboratory of the Italian Space Agency (ASI), while Cagliari and Kryoneri are two locations that are not associated with any specific infrastructure. We expressly selected a few sites in which we expected a bad seeing\footnote{By 'bad seeing' we mean substantially worse than in a good astronomical site.} to carry out this analysis.
\end{enumerate}

In Section \ref{def} we provide the definitions of the forecast timescale and of precasting and prediction by persistence methods. In Section \ref{obs} we present the observations used in this paper. In Section \ref{ana}, we present the analysis and in Section \ref{disc}, we discuss motivation (3). Finally, in Section \ref{conc}, we synthesise the conclusions of the paper.
   
\section{Models and forecast timescale}
\label{def}

\begin{figure}
\centering
\includegraphics[width=\hsize]{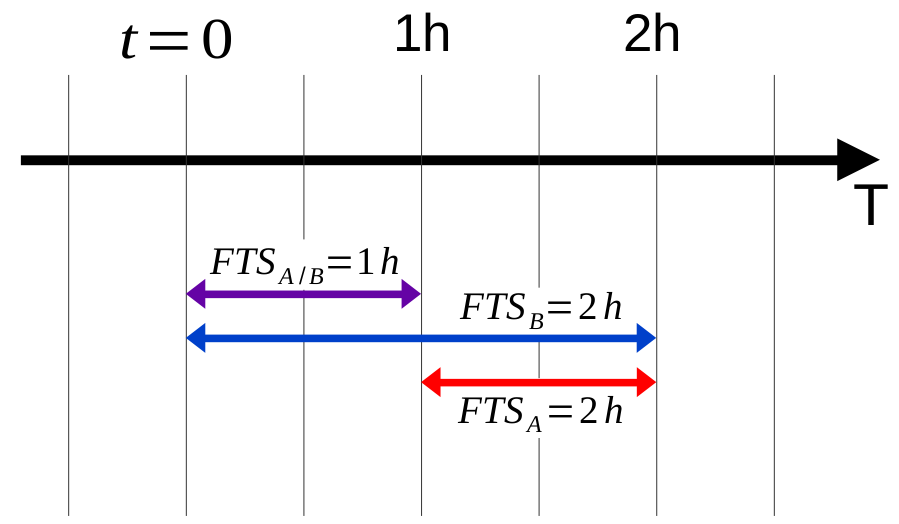}
\caption{Schematised representation of cases (A) and (B) of FTS related to 1h and 2h. The arrows indicate the start and end of each interval as defined in the paper. The present time is indicated with t=0 as described in the paper. } 
\label{FTS_defintion}
\end{figure}

In this section, we give the definition of the different models as well as the definition of a few concepts treated in the paper: 

\begin{itemize}

\item Prediction by persistence (PP) is a reference method useful when we are forecasting the temporal evolution of a parameter. It is a method that assumes that the variable of interest remains constant during the interval of time that we are forecasting. For instance, if the seeing has a value $\see$($t_{0}$) at time $t_0$, then the forecast simply predicts that the seeing remains equal to $\see$($t_{0}$) over the entire period of time that we are considering. In this paper we study whether, instead of taking $t_0$, we may obtain a more representative estimate by assuming the average of $\varepsilon$ on the last  $\Delta(t)$ minutes before the time $t_{0}$ where  $\Delta(t)$ can be 5~minutes up to 1~hour. We define the optimal value of  $\Delta(t)$ as the one that minimises the   root mean square error (RMSE) with respect to the observations. To eliminate the very high temporal frequencies that are not relevant in our application,  after the calculation of the average in the $\Delta(t)$ observations are processed with a centred 1h moving average.\footnote {The necessity of a moving average of the data is the conclusion of a feasibility study conducted in the past for ESO in the project (MOSE). The main reason is related to the fact that when we talk about `seeing forecast` we are referring to the forecast of a trend, therefore we are not referring to the high frequencies.}

\item Precasting\footnote{The term 'precasting' is used by European Southern Observatory to indicate the quantity defined in this paper with the term PC. We use the same term for consistency with further work/studies done using this term.} (PC) is a reference method useful when we are forecasting the average of a parameter within an interval of time. It consists of calculating the mean of the parameter in a window of  $\Delta(t^{*})$ minutes before $t_0$, and assuming it will also be the average value of the interval of interest. At present, the most critical and useful time interval for science operations is the upcoming 1h. In this paper we study which is the optimal value of  $\Delta(t^{*})$, in the same manner as for the PP.  Even if in principle $\Delta(t)$ is not necessarily equal to $\Delta(t^{*})$, later on we will see that the optimal values of $\Delta(t)$ and $\Delta(t^{*})$ are very similar.

 The calculation of the PC and PP forecasts are similar but they represent different things, as we have defined above. We have a scalar in the PC case and a vector in the PP case. Another difference is how we treat  observations in comparison with forecasts, as in the case of PC we do not perform a moving average of  observations as we are estimating an individual value and we are not forecasting the temporal evolution of a parameter. 

 In both cases (PC and PP) we recalculate the forecasts every 10 minutes. The reason of this choice is simply that this corresponds to the technical specifications of a project we are carrying out for ESO. It was therefore practical for us to adopt the same conditions.

\item The concept of forecast timescale (FTS) has been extensively described in the introduction in \citep{masciadri2023} where a precise definition is provided from a general point of view and for some specific cases (long and short forecast timescales). We report here the core of the definition and we refer the reader to \citep{masciadri2023} for supplementary information. The FTS is the interval between the time at which the initialisation forecast is calculated ($T_{fc}$) and the time to which the forecast refers. Assuming that we want to forecast the temporal evolution in the interval $T_{ini}$ to $T_{end}$, the FTS is included in the [($T_{ini}$~-~$T_{fc}$), ($T_{end}$~-~$T_{fc}$)].

In the case of the PC and the PP, the forecasts are computed from real-time observations, and the forecast is therefore calculated with respect to the present time, t=0. 
That means that $T_{fc}$ = 0, therefore the interval of forecast becomes [$T_{ini}$, $T_{end}$].  In case the FTS are multiples of 1 hour (i.e. Xh) the intervals correspond to [t$_{0}$ + X~-~1, t$_{0}$ + X]. This method has been proposed by \citep{masciadri2023} for the autoregression (AR) method. 
For the forecast at 1h we have, for example, $T_{ini}$ = 0 and $T_{end}$ = 1h. 
For the forecast at 2h, we have $T_{ini}$ = 1 and $T_{end}$ = 2 with respect to the preset time = 0. A 2h forecast, therefore, applies to all values within the [1h, 2h] range. We call this FTS case (A) hereafter.  

In this paper we additionally consider an alternative definition that we call case (B), in which $T_{ini}$ is always equal to 0 and $T_{end}$ is equal to the end of the interval of interest. In the previous example of a forecast at 2h we now have $T_{ini}$ = 0 and $T_{end}$ = 2h. Figure \ref{FTS_defintion} reports the schematic representation of the FTS for cases (A) and (B). We note that, by definition, the FTS at 1-hour is the same for case (A) and case (B). Case (B) has the characteristics to be more flexible as we can define in an easier way whatever interval of time, even if such a criterion gives more weight to the first part of the interval while case (A) gives more weight to the last part of the interval, and it should be preferable when we talk about a forecast as the forecast decreases in quality with the time.  It is important to note that, by construction, case (B) provides better performances as soon as the FTS increases, but this does not mean that it is a better criterion. The important  thing is to use the same criterion between the reference and forecast methods. We use these two different definitions to highlight the importance of the criteria used to define the FTS.  Method (A) (also used in \citet{masciadri2020,turchi2020,masciadri2023} but for the AR approach) is preferable as it is more representative if there is no specific reason to do differently.

\end{itemize}

\section{Observations}
\label{obs}
For this study, we used observations of optical turbulence and atmospheric parameters relevant for ground-based astronomy at seven sites: (1) Cerro Paranal (Chile), the site of the VLT, (2) Mt. Graham (US), the site of the LBT, (3) Roque de los Muchachos Observatory (La Palma, Canary Islands), site of the Telescopio Nazionale Galileo (TNG)\footnote{We used the  Differential Image Motion Monitor (DIMM) of the TNG.}, (4) La Silla (Chile), an ESO site,  location of the New Technology Telescope (NTT), (5) the Matera Space Centre Laboratory of the ASI close to Matera (Italy), (6) Cagliari (Italy), and (7) Kryoneri (Greece). The main analysis focuses on the first two sites. The last five sites are used for a specific analysis presented in Section \ref{disc}.

For Paranal, we used observations stored in a public repository of ESO, the so-called ESO’s astronomical Ambient Conditions database\footnote{\href{https://archive.eso.org/cms/eso-data/ambient-conditions/paranal-ambient-query-forms.html}{https://archive.eso.org/cms/eso-data/ambient-conditions/paranal-ambient-query-forms.html}}. Astroclimatic parameter measurements start from April 2016. A DIMM provides measurements of the seeing, i.e. the integrated turbulence along the line of sight corrected for zenith angle. DIMM measurements are based on the variance of the angle-of-arrival fluctuations of a stellar wavefront passing through two sub-apertures of a small telescope, typically a class 30~cm \citep{sarazin1990}. The other astroclimatic parameters, the isoplanatic angle ($\theta_0$), wavefront coherence time ($\tau_0$), and the GLF are measured with a MASS-DIMM. The Multi-Aperture Scintillation Sensor (MASS) is a vertical profiler and was introduced in the astronomical context by \citep{kornilov2003}, using a slightly modified version of the technique proposed by \citep{ochs1976}. It reconstructs turbulence stratification (\CN2) in six layers distributed within 20 km above the ground, starting from scintillation indices (four normal and six differential indices) calculated from scintillation maps produced by single stars. The MASS is blind to turbulence close to the surface and the MASS–DIMM instrument \citep{kornilov2014} combines the properties of the DIMM and the MASS allowing reconstructions of all astroclimatic parameters for 20~km above the ground. The GLF is defined as the ratio between the turbulence close to the ground and the total turbulence developed in the whole atmosphere, i.e.  $\sim$ 20~km. We can recover the contribution of the turbulence associated with the ground by subtracting the MASS contribution from the DIMM contribution

\begin{equation}
GLF = \frac{J_{GL}}{J_{GL}+J_{MASS}},
\label{eq1_GLF}
\end{equation}
\noindent
where the generic contribution J is
\begin{equation}
J = \int_{0}^{\infty }C_{N}^{2}dh,
\label{eq2_GLF}
\end{equation}

\noindent
and J$_{GL}$ is the J retrieved from the DIMM minus J$_{MASS}$. We refer to Section 4 in \citet{masciadri2023} for a more extended discussion on GLF. The temporal sampling of the DIMM and MASS-DIMM measurements is not constant, with an average of 1.5~min. 

Atmospheric parameters (WS, RH, T at 30~m above ground level (AGL), PWV) are measured with a Vaisala meteorological station, which was installed in Paranal in 1984 and upgraded in 1998 \citep{eso_amb_cond_2016}.  All measurements in the database are resampled by ESO on a timescale of one minute. The PWV is measured with three microwave radiometers called LHATPRO \citep{querel2014} at three different locations on top of the Paranal plateau. Statistics of PWV observations are much more important (almost a factor of two) during the day than during the night, as in the former case the instrument runs continuously along the zenith direction, while at night, since 2021, radiometers are forced to point at different lines of sight during the night as they are also used to collect measurements useful to calibrate telluric lines.  

At Mt. Graham, the seeing measurements were done with a DIMM that is located inside the dome of the LBT. This implies that measurements also include part of the dome seeing, but this solution has the advantage of providing a turbulence estimate closer to what is seen by the telescope when an astronomical object is imaged on the detector. We therefore always used DIMM measurements as a reference, while keeping in mind that the typical estimates are probably slightly overestimated by the dome contribution, but this fits well with our requirements. The temporal sampling of DIMM measurements is not constant, with an average of $\sim$ 6~sec. At Mt. Graham there is no vertical profiler running routinely \footnote{The implementation of a Ring-Image Next Generation Turbulence Sensor (RINGSS) is ongoing. Data will be available as soon as the instrument is in commission.}; therefore, other astroclimatic parameters than the seeing cannot be observed routinely. 
Atmospheric parameters (WS, RH, T) are obtained from a weather mast placed 5~m above the roof of the dome, at its rear. A second mast placed 3~m above the front of the roof also measures the wind speed. Tests have shown optimal results when taking measurements from the front mast when the wind comes from the front ($\pm 30$°), otherwise from the back mast. For more details, refer to \citet{turchi2017}. The temporal sampling for all atmospheric parameters is 1~sec. 

For both astroclimatic and atmospheric parameters above the two sites of Mt. Graham and Cerro Paranal, we used eight years of data, from 2017/01/01 to 2024/12/31.
 Above Paranal there are no measurements between March 2020 and September 2020 due to COVID-19 shutdowns. MASS experienced a technical issue in 2017, so GLF, $\tau_0$, and $\theta_0$ were not measured between February and June 2017. Only a negligible number of dates lacked sufficient meteorological data. Temporal sampling of seeing measurements is irregular, with a median of 79 s. 
Above Mt. Graham, the LBT is closed during July and August due to the monsoon season. The temporal sampling of seeing measurements is irregular, with a median of 5s. 

The atmospheric parameters for both the VLT and the LBT were measured during the full 24 hours and were segmented into night and day subsets. 'Night' is defined as the period during which seeing measurements are available, or otherwise from 19:00 to 06:00 local time (LT), which corresponds to UTC-4 for VLT and UTC-7 for LBT.

At Roque de los Muchachos Observatory in La Palma (Canary Islands), we considered the seeing measured by the DIMM of the TNG, in operation since 2011 \citep{molinari2012,deguturbai2013}. 
The temporal sampling of the measurements is not constant, with an average of 55~s. We used eight years of data in the same range [2017/01/01, 2024/12/31] used for VLT and LBT. 

At La Silla Observatory, the seeing is measured with a DIMM. Measurements are stored in a public repository of ESO\footnote{\href{https://www.eso.org/sci/facilities/lasilla/astclim/seeing.html}{https://www.eso.org/sci/facilities/lasilla/astclim/seeing.html}}. This DIMM was decomissioned in 2024 to be replaced by a RINGSS. Therefore, we limited the usage of the measurements to seven years [2017/01/01, 2023/12/31] instead of eight. The temporal sampling of the measurements is not constant, with a median of 62s. 

At the Matera Space Lab, a non-astronomical site, the seeing is measured with a Cyclope \citep{dicecco2023}. Cyclope is an instrument that quantifies the seeing by looking at the polar star, but it does not perform a differential measurement. For this reason, the instrument must be fixed to the ground to avoid vibrations, which are eliminated by the differential measurement of the DIMM. It has been developed by the company Alcor System\footnote{\href{www.alcor-system.com/new}{www.alcor-system.com/new}}. 
A few comparisons of a few hours with a DIMM in strong turbulence conditions are available on the Cyclope documentation and the Alcor web page. We verified that the instrument can reach seeing values as low as 0.5$\arcsec$ therefore we should not have biases on a specific range of the spectrum of the turbulence. The temporal sampling of the measurements was not constant with an average of 55~s. We used one year of data [2022/01/01, 2022/12/31] as this was the unique homogeneous period of measurements. 

At the sites of Cagliari and Kryoneri, the seeing was measured with a C-DIMM developed by the company Miratlas\footnote{\href{miratlas.com}{miratlas.com}}. The C-DIMM is a slightly modified version of a DIMM as it is based on the same principle but uses two independent telescopes with a pupil size of 5~cm. 
The temporal sampling of both sites was regular and equal to 60s. We used two years of data [2024/01/01,2025/12/31].  

We highlight that the cases of Matera, Cagliari, and Kryoneri are statistically less representative than the others, as the sample covers one or two years instead of eight years. That said, we expressly wanted to consider a few sites characterised by a worse seeing than typically measured above astronomical sites of top-class telescopes. In Appendix \ref{ann_a} the histograms with the statistics of the different sites are given. 

\section{Analysis}
\label{ana}

\begin{figure}
\centering
\includegraphics[width=\hsize]{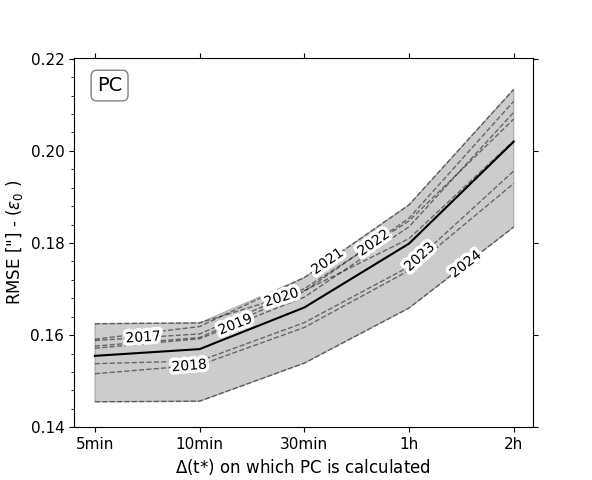}
\caption{RMSE between observations and PC above  Cerro Paranal for a FTS = 1h as a function of the interval of time  $\Delta(t^*)$ over which the precasting is calculated. The bold black line is the RMSE obtained with the whole sample of 8 years. The dashed lines report RMSE obtained in each individual year. The grey area represents the uncertainty.} 
\label{criterium}
\end{figure}

\begin{figure}
\centering
\includegraphics[width=\hsize]{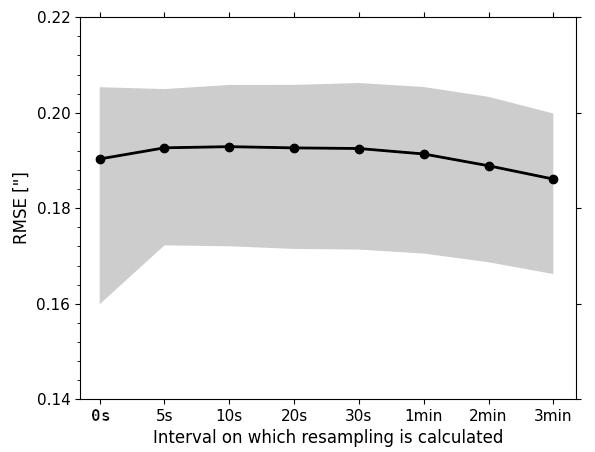}
\caption{RMSE between observations and PC for a FTS =1h as a function of different intervals of time on which the measurements are resampled. The calculation is performed for the seeing at Mt. Graham. The dataset spans the period [2017-2024].}
\label{resampling}
\end{figure}

\begin{figure*}
  \centering
  \includegraphics[width=\hsize]{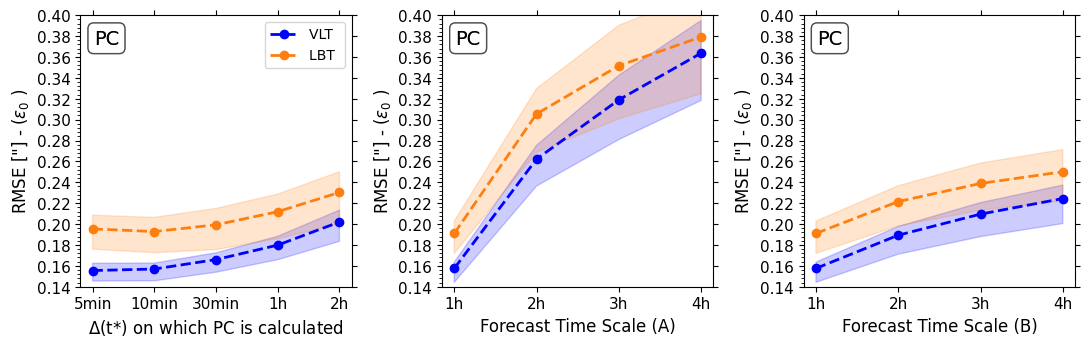}
  \caption{Seeing: RMSE between observations and PC above Paranal (blue line) and Mt. Graham (orange line) using the dataset belonging to the range of 8 years [2017,2024]. Left: RMSE as a function of the interval of time  $\Delta(t^*)$  on which the PC is calculated for FTS = 1h. Centre: RMSE as a function of the forecast timescales~-~case (A). Right: RMSE as a function of the forecast timescales for case (B). In the last two columns for each parameter we use the optimal  $\Delta(t^*)$ (10min) retrieved from the first column. See text in Section \protect\ref{ana}.}
  \label{seeing_PC}
\end{figure*}

\begin{figure*}
  \centering
  \includegraphics[width=\hsize]{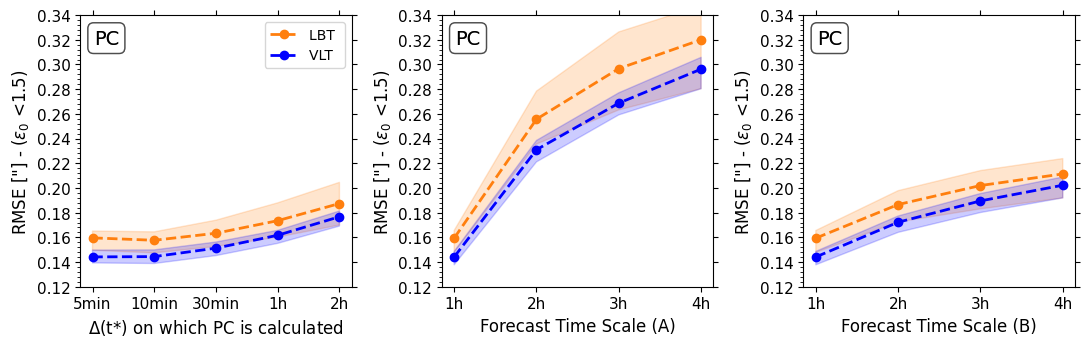}
  \caption{Same as Fig.\protect\ref{seeing_PC} but for the $\varepsilon$ < 1.5 $\arcsec$.  That means that when we calculate the RMSE, we consider only the couples in which the element related to observations is below 1.5 $\arcsec$. The optimal  $\Delta(t^*)$  is 10 minutes.  }
  \label{seeing_15_PC}
\end{figure*}

\begin{figure*}
  \centering
  \includegraphics[width=\hsize]{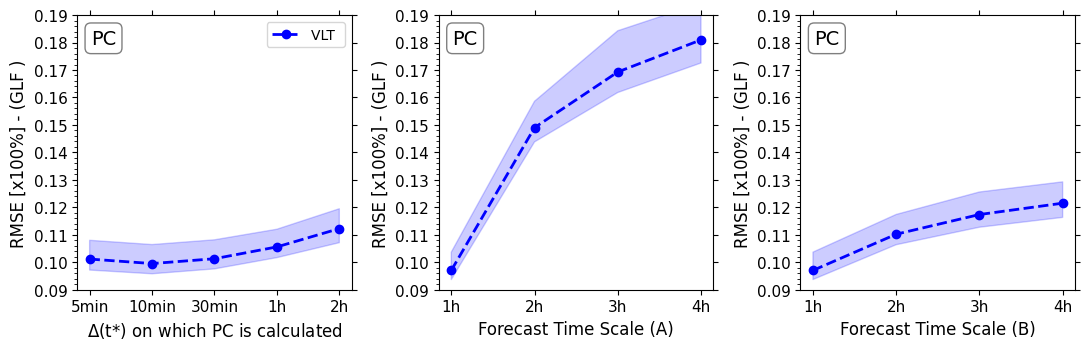}
  \caption{Same as Fig.\protect\ref{seeing_PC} but for the GLF. The optimal  $\Delta(t^*)$  is 10 minutes.}
  \label{GLF_PC}
\end{figure*}

\begin{figure*}
  \centering
  \includegraphics[width=\hsize]{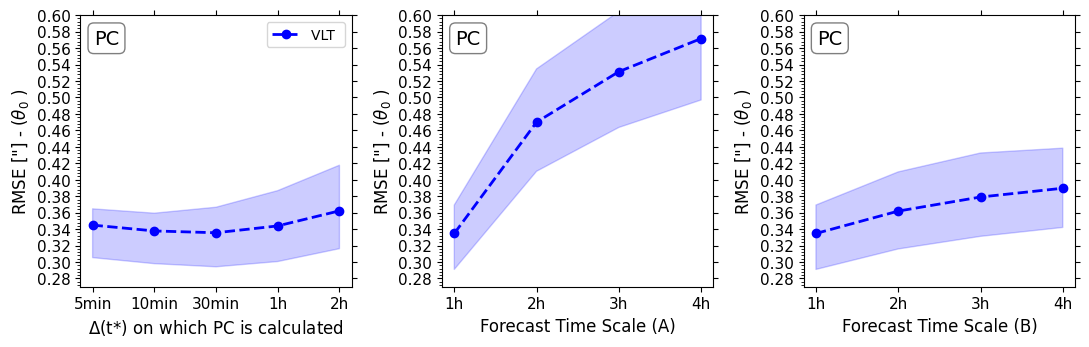}
  \caption{Same as Fig.\protect\ref{seeing_PC} but for the  isoplanatic angle. The optimal  $\Delta(t^*)$  is 30 minutes.}
  \label{Theta_PC}
\end{figure*}

\begin{figure*}
  \centering
  \includegraphics[width=\hsize]{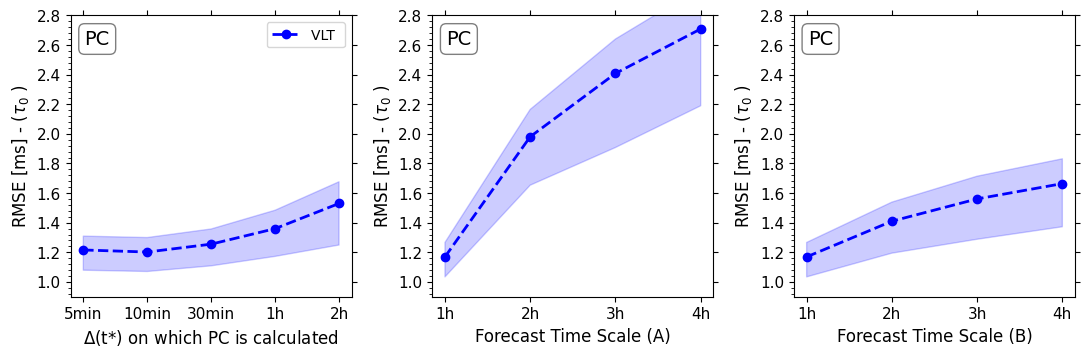}
  \caption{Same as Fig.\protect\ref{seeing_PC} but for the  wavefront coherence time. The optimal  $\Delta(t^*)$  is 10 minutes.}
  \label{Tau_PC} 
\end{figure*}

\begin{figure*}
\centering
\includegraphics[width=\hsize]{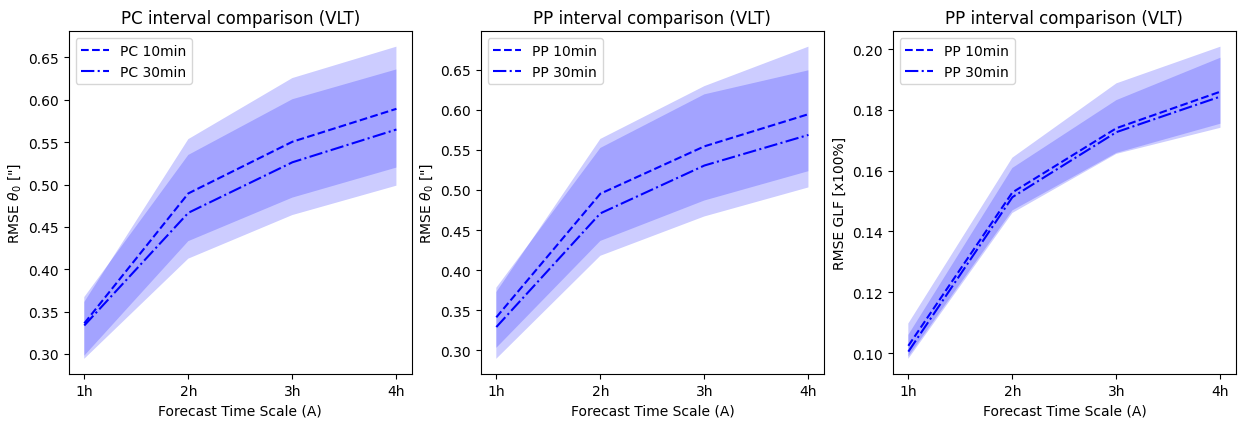}
\caption{RMSE calculated for the $\theta_{0}$ and GLF above the VLT as a function of the FTS in the case of PC and PP assuming a  $\Delta(t^*)$ and $\Delta(t)$ of 10 and 30 minutes. } 
\label{theta0}
\end{figure*}

\begin{figure*}
  \centering
  \includegraphics[width=\hsize]{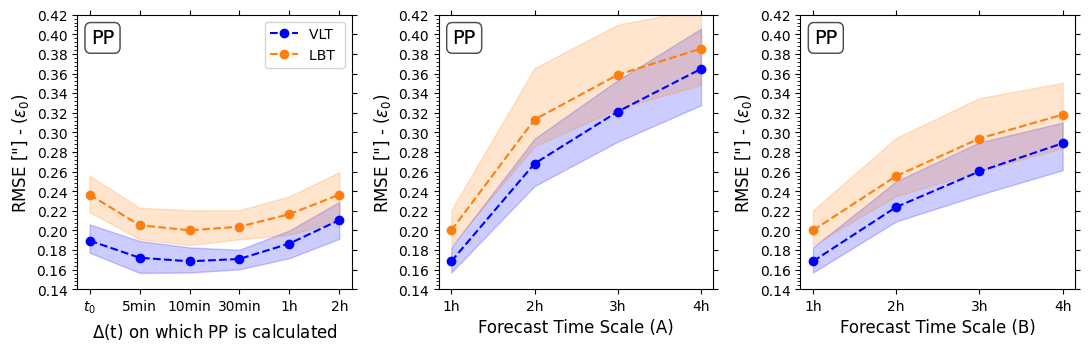}
  \caption{Seeing: RMSE between observations and prediction by PP above Paranal (blue line) and Mt. Graham (orange line) using the dataset belonging to the range of 8 years [2017,2024]. Left: RMSE as a function of the interval of time $\Delta$t on which the PP is calculated for FTS = 1h. Centre: RMSE as a function of the forecast timescales~-~case (A). Right: RMSE as a function of the forecast timescales for case (B). In the last two columns for each parameter we use the optimal $\Delta$t (10min) retrieved from the first column. See text in Section \protect\ref{ana}.}
  \label{seeing_PP}
\end{figure*}

\begin{figure*}
  \centering
  \includegraphics[width=\hsize]{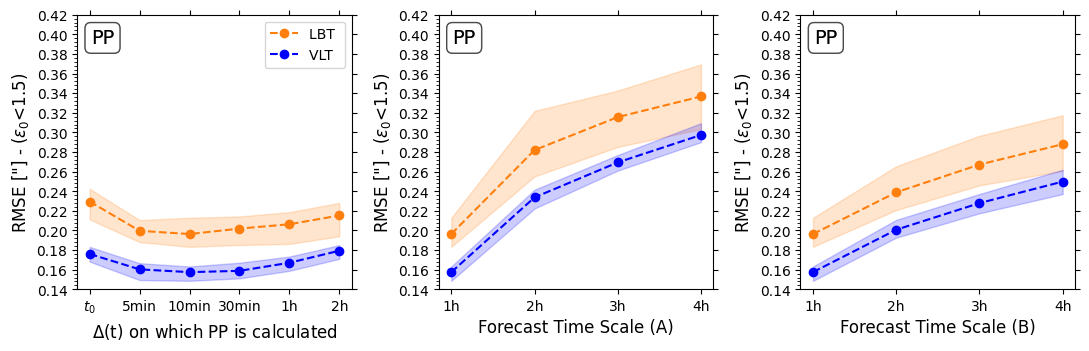}
  \caption{Same as Fig.\protect\ref{seeing_PP} but for the $\varepsilon$ < 1.5 $\arcsec$. The optimal  $\Delta(t)$ is 10~min.}
  \label{seeing_15_PP}
\end{figure*}

\begin{figure*}
  \centering
  \includegraphics[width=\hsize]{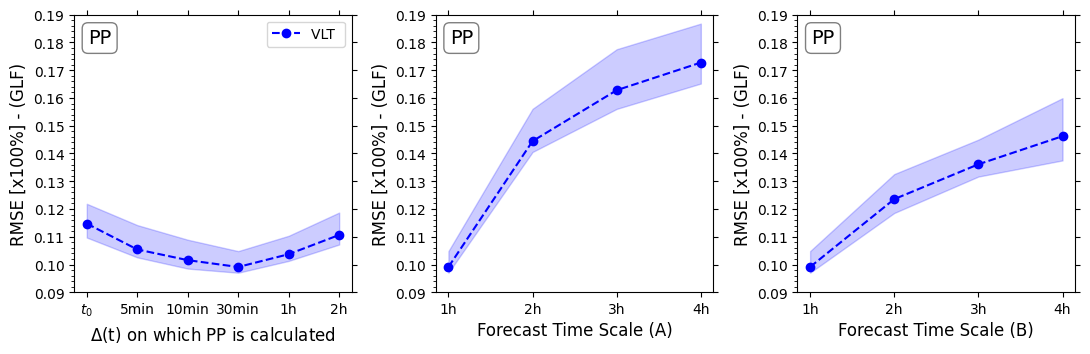}
  \caption{Same as Fig.\protect\ref{seeing_PP} but for the GLF. The optimal  $\Delta(t)$ is 30~min.}
  \label{GLF_PP}
\end{figure*}

\begin{figure*}
  \centering
  \includegraphics[width=\hsize]{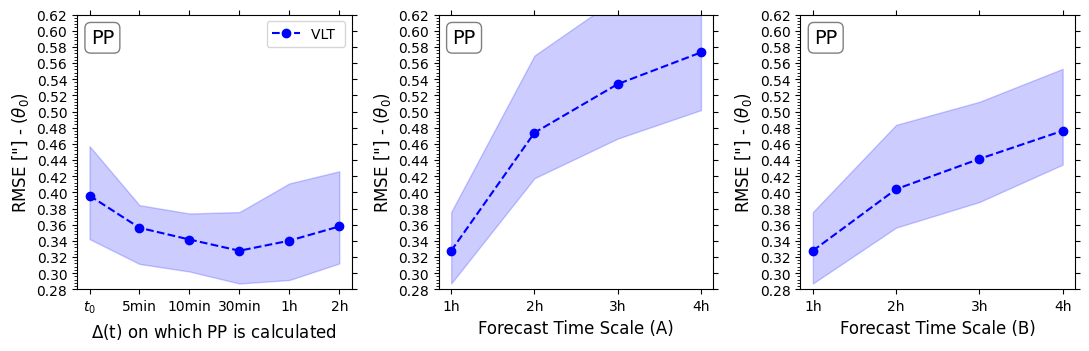}
  \caption{Same as Fig.\protect\ref{seeing_PP} but for the  isoplanatic angle. The optimal  $\Delta(t)$ is 30~min}.
  \label{Theta_PP}
\end{figure*}

\begin{figure*}
  \centering
  \includegraphics[width=\hsize]{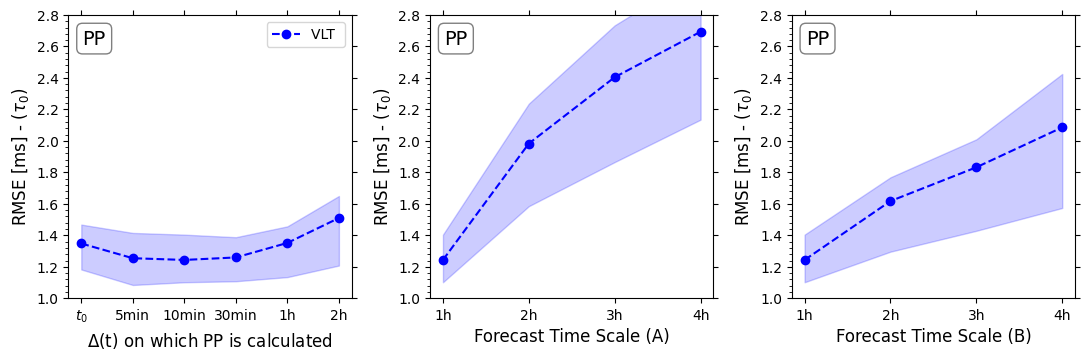}
  \caption{Same as Fig.\protect\ref{seeing_PP} but for the  wavefront coherence time. The optimal  $\Delta(t)$ is 10~min.}
  \label{Tau_PP} 
\end{figure*}

\begin{table*}
\centering
\caption{RMSE between observations and PC and PP. }
\ra{1.3}
\begin{tabular}{@{}rcrrrcrrr@{}}\toprule
Parameters & \multicolumn{3}{c}{PP} & \phantom{abc}& \multicolumn{3}{c}{PC}\\
\cmidrule{2-4} \cmidrule{6-8}
& 1h & 2h (FTS A) & 2h (FTS B) && 1h & 2h (FTS A) & 2h (FTS B)\\ \midrule
\bf{VLT} \\
$\varepsilon$ [``] &  0.17 & 0.27 & 0.22 && 0.16 & 0.26 & 0.19 \\
$\varepsilon \le 1.5$ [``] &  0.16 &  0.23 &  0.20 && 0.14 & 0.23 & 0.17 \\
$\tau_0$ [ms] &  1.24 &  1.96 &  1.60 && 1.17 & 1.98 & 1.41 \\
$\theta_0$[``] & 0.33 &  0.48 & 0.40 && 0.34 & 0.47 & 0.37   \\
$GLF$ [\%] & 10.0 & 14.6 & 12.3 && 09.7 & 14.6 & 11.0 \\
$PWV_{night}$ [mm] & 0.24 & 0.47 & 0.36 && 0.21 & 0.46 & 0.31 \\
$PWV_{day}$ [mm]  & 0.27 &  0.53 & 0.41 && 0.24 & 0.52 & 0.35 \\
$T_{night}$ [°C] & 0.33 & 0.61 & 0.48 && 0.30 & 0.59 & 0.41 \\
$T_{day}$ [°C] &  0.41 &  0.82 &  0.64 && 0.36 & 0.81 & 0.54 \\
$RH_{night}$ [\%] & 2.31 & 4.21 &  3.42 && 2.11 & 4.19 & 2.85 \\
$RH_{day}$ [\%] &  2.31 & 4.21 &  3.42 && 2.12 & 4.27 & 2.91 \\
$WS_{night}$ [$ms^{-1}$] & 0.94 & 1.73 &  1.39 && 0.90 & 1.72 & 1.19   \\
$WS_{day}$ [$ms^{-1}$] & 0.93 &  1.73 &  1.39 && 0.89 & 1.73 & 1.20 \\

\bf{LBT} \\
$\varepsilon$ [``] &   0.20 &  0.31 &  0.25 && 0.19 & 0.31 & 0.22\\
$\varepsilon$ < 1.5 [``] &   0.20 &  0.28 &  0.24 && 0.16 & 0.26 & 0.19 \\
$T_{night}$ [°C] & 0.41 & 0.77 & 0.61 && 0.39 & 0.78 & 0.52 \\
$T_{day}$ [°C] &  0.55 &  1.07 & 0.84 && 0.51 & 1.06 & 0.72\\
$RH_{night}$[\%] & 4.49 & 8.26 &  6.65 && 4.09 & 8.31 & 5.55 \\
$RH_{day}$[\%] & 4.47 &  8.24 &  6.62 && 4.10 & 8.26 & 5.66 \\
$WS_{night}$ [$ms^{-1}$] & 1.53 &  2.25 &  1.93 && 1.45 & 2.26 & 1.67 \\
$WS_{day}$[$ms^{-1}$] &  1.32 &  2.02 &  1.72 && 1.25 & 2.01 & 1.49 \\

\bottomrule
\end{tabular}
\label{table_results}
\tablefoot{Values of RMSE for FTS = 1h and 2h in case (A) and (B) are extracted from Fig.\protect\ref{seeing_PC} to Fig.\protect\ref{Theta_PP}. RMSE at FTS=1h for case (A) and case (B) are the same by construction. This table also includes values related to the analysis of the atmospheric parameters treated in Annex \protect\ref{ann_d}.}
\end{table*}

Here we describe the analysis performed to address the questions outlined in Section\ref{intro}. 
At points (1) and (2) we present the key elements of the preliminary data processing that was implemented:\\ \\

(1) The goal of the analysis is to quantify the accuracy of the two methods (PC and PP) using the RMSE between observations and forecasts as a figure of merit. This metric has been widely used in the past in many studies in the literature \citep{lascaux2013, lascaux2015,turchi2017,turchi2020,masciadri2020,masciadri2023} and it appears to be the most appropriate choice to assess the accuracy of the methods considered. In the analysis presented in this paper we consider the RMSE value, which represents the associated uncertainty (or equivalently, the accuracy) plus a sort of 'dispersion'. This dispersion is not the $\sigma$ related to BIAS and RMSE as reported in Eq.1, 2, and 3 of \citep{masciadri2023}. 
Figure \ref{criterium} shows the criteria used in this paper to define this dispersion. The thick black line indicates the RMSE computed on the entire dataset spanning eight years, while the grey band is the dispersion that is the space covered by the RMSE calculated for each individual year.  We resample our data into 10-minutes bins before calculating the RMSE.

 The RMSE between forecast and observations for the PC case is
\begin{equation}
RMSE_{PC}=\sqrt{\frac{1}{M}\sum_{k=1}^{M}(F_k-O_k)^2},
\label{eqRMSEPC}
\end{equation}

 where $F_k$ are the forecasts of the average for the upcoming FTS calculated at timestamp $k$, $O_k$ is the actual average observed over the same FTS, and $M$ is the total number of timestamps in the evaluation period.

 The RMSE between forecast and observations for the PP case is
\begin{equation}
RMSE_{PP}=\sqrt{\frac{1}{N_{tot}}\sum_{i=1}^{N}\sum_{k=1}^{K_i}(F_{i}-O_{i,k})^2},
\label{eqRMSEPP}
\end{equation}

 where $F_{i}$ is the value of the $i^{th}$ forecast interval and this value is held constant for all  $k$ sampling elements within the FTS. $O_{i,k}$ denotes the $k^{th}$ actual observation related to the $i^{th}$ forecast interval, $K_i$ is the number of sampling elements within the FTS during the $i^{the}$ forecast,  $N$ is the total number of forecast intervals, and $N_{tot} = \sum_{i=1}^{N} K_i$ represents the total number of all evaluated observation sampling points. We highlight that for the PP case, the 1h moving average described in Section 2 is done before performing the resampling of 10 minutes. 

(2) A further important issue concerns the typical sampling frequency of the raw data. The DIMM-VLT observations are sampled at an irregular frequency ranging from about 1~min to around 1.5~min whereas the DIMM-LBT observations are sampled every few seconds. Figure \ref{resampling} shows the RMSE obtained from DIMM-LBT data after applying different resampling intervals, from 5~s up to 3~min. The results indicate that the RMSE remains basically constant between 5~s and 1.5~min, and it begins to decrease above 2~min. This implies that the statistical properties of the turbulence are preserved when the data are sampled at a frequency of up to around 1.5-2~min. Therefore, data from the two sites can be considered homogeneous and comparable. As anticipated in Section \ref{intro}, here we extensively treat the cases of DIMM-LBT and DIMM-VLT. The other sites (TNG, La Silla, Matera, Cagliari, Kryoneri) are only considered for a specific analysis that is discussed later in Section \ref{disc}. For completeness, we reiterate that the five other sites have a temporal sampling close to one minute and therefore can be considered homogeneous with the cases of Paranal and Mt. Graham (Fig. \ref{resampling}).\\ 

From Fig.\ref{seeing_PC} to Fig.\ref{Theta_PC} we report the RMSE between the observations and the PC related to all the main available astroclimatic parameters above Paranal (blue line) and Mt. Graham (orange line) calculated in the dataset in the range  [2017, 2024]. The astroclimatic parameters are: the seeing ($\varepsilon$), the $\varepsilon$ < 1.5 arcsec, the wavefront coherence time ($\tau_{0}$), the GLF, and the isoplanatic angle ($\theta_{0}$) (see in Appendix \ref{ann_b} the analytical expression of these parameters). The study of $\varepsilon$ < 1.5 arcsec is relevant, as, in general, the forecast performance decreases with increasing seeing. However, it is less interesting to forecast seeing values larger than 1.5 arcsec, as adaptive optics cannot be used under such atmospheric conditions. This threshold can be considered as a conservative range in which to guarantee the desired performances. We note that a few parameters cannot be calculated in the case of Mt. Graham because of the lack of instruments measuring the specific parameters. In particular, we cannot measure $\tau_{0}$, $\theta_{0}$, and the GLF as a vertical profiler of the optical turbulence is not yet operational  at LBT (see Section \ref{obs}) and we cannot measure PWV as a radiometer is missing. In the first column of Fig.\ref{seeing_PC}~-~Fig.\ref{Theta_PC} is shown the RMSE between the observations and the PC (calculated for a FTS = 1h) as a function of the interval of time  $\Delta(t^*)$ on which is calculated the average of the signal (see the PC definition in Section \ref{def}). In the central column, the RMSE between observations and the PC is shown for case (A), and, in the right column, the same thing is shown for case (B). In the first column, we extract the optimal value of  $\Delta(t^*)$, which is the one that provides the minimum RMSE. We observe that for $\theta_{0}$ the optimal   $\Delta(t^*)$ = 30 min. For all other astroclimatic parameters  $\Delta(t^*)$ = 10 min.  In some cases, the value of 10 and 5 minutes is basically the same (the difference is absolutely negligible). For those cases, we decided to use 10 minutes to homogenize the data treatment, as this covers the majority of cases and allows a calculation on a larger statistics of points than the case of 5 minutes. In the central and right columns, the optimal  $\Delta(t^*)$ for each parameter has been used in the forecast calculations.  As expected and as explained in Section 2, the RMSE in case (A) provides worse results than in case B for a FTS > 1h. The deterioration of RMSE is obviously sharper in case (A) than in case (B), since the second is more influenced by the good values close to t$_{0}$ while case (A) is more influenced by the values far from t$_{0}$. 

In the left panel of Fig.\ref{theta0}, we show the RMSE of $\theta_{0}$ between observations and PC scales with different FTS using a different value of  $\Delta(t^*)$. The difference remains within a maximum of 3.5\% for a  $\Delta(t^*)$ = 1h for both cases (A) and (B), which is a negligible quantity. We can therefore reasonably use the same 10-min interval for all the parameters.  Figure \ref{theta0} central and right panels are discussed later in relation to the PP. Finally, we reiterate that, as mentioned in Section \ref{def}, the RMSE for a FTS = 1h in case (A) and (B) is the same by construction. \\

From Fig.\ref{seeing_PP} to Fig.\ref{Theta_PP} we report the RMSE between the observations and the PP related to all the main astroclimatic parameters above Paranal and Mt. Graham calculated on the [2017, 2024] dataset. The astroclimatic parameters are treated in the same way as for the PC case. From Fig.\ref{seeing_PP} to Fig.\ref{Theta_PP} (column 1), we deduce that the values of $\Delta(t)$ that provide the optimal RMSE between observations and PP at FTS=1h  are very similar to those identified for  $\Delta(t^*)$ in the PC case.  For the PP case, we observe that $\theta_{0}$ and GLF show an optimal $\Delta(t)$ of 30 min instead of 10 min. The impact on the RMSE is negligible as in the PC case.

Table \ref{table_results} reports, for each parameter and at both astronomical sites, the RMSE values  between observations and the corresponding forecast methods (PC and PP) for the FTS = 1h and 2h that are the most relevant for the science operations of ground-based astronomy, particularly for VLT and LBT. The cases (A) and (B) are considered in Table \ref{table_results}. The values in Table \ref{table_results} are the thresholds to be compared with any considered forecast method, regardless of the techniques for the forecast of the astroclimatic and atmospheric parameters above the LBT and VLT sites. This paper therefore reports key information to quantify the goodness and utility of a forecast method above the mentioned sites. In Annex \ref{ann_d}, the results of the same analysis related to atmospheric parameters are shown.

Looking at the second column in Fig.\ref{seeing_PC}-Fig.\ref{seeing_15_PC} and Fig.\ref{seeing_PP}-Fig.\ref{seeing_15_PP}, we observe that PC and PP performances are better at Paranal than at Mt. Graham for the seeing for all the forecast timescales\footnote{The seeing $\varepsilon$ is the unique astroclimatic parameter for which a comparison above LBT and VLT is allowed for the reasons we have already explained.}. As a first guess, one might assume that the explanation for what we observe is that turbulence at VLT is more stable than at LBT, but this is merely a hypothesis, and it would be interesting to identify a physical explanation for such a difference. This is the subject of Section \ref{disc}.  As anticipated in Section\ref{def}, for FTS > 1h, case (A) provides worse performances than case (B) for construction. We recommend using case (A) instead of (B) if there are no particular reasons to choose (B) as it is more representative. We remind the reader that the goal of this paper is to provide performances of a couple of simple reference methods; therefore, to appreciate if a more complex method and/or model provides useful information, one has to compare RMSE performances of that model with those of the references using the same configuration, i.e. the same case (A) or (B). Looking at Fig.\ref{seeing_PC}-Fig.\ref{Tau_PP}, we can observe that forecast performances decrease with the increase of FTS as we might expect. The goal is therefore to overcome these thresholds at the different FTS using different methods and/or models employing different strategies. 

\section{Discussion}
\label{disc}

\begin{figure*}
\centering
\includegraphics[width=0.9\hsize]{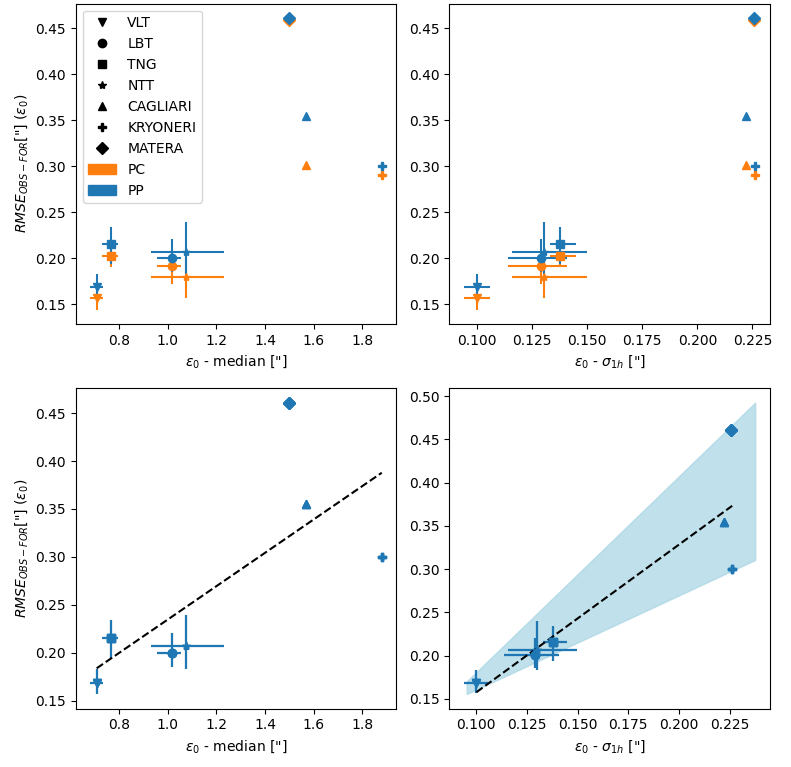}
\caption{Top: Scattering plot between the RMSE$_{OBS-FOR}$ of the seeing with the median (left) and median hourly standard (right) of the historical data for PP and PC. FTS = 1h, $\Delta t$ = 10min. Bottom: For the PC case, the dashed black line indicates the regression line between the RMSE$_{OBS-FOR}$ and the median (left) and the median  $\sigma_{1h}$ (right). The case for PP is equivalent. }
     \label{RMSE_vs_mean_sigma}
\end{figure*}

In this section we investigate whether it is possible to find a correlation between the statistical properties of the seeing and RMSE$_{OBS-FOR}$ between the observations and PC and the observations and PP that we found in Section \ref{ana}. We are looking for some law that allows us to retrieve values of the reference thresholds of RMSE$_{OBS-FOR}$ for the seeing above other sites. The challenge is to investigate whether threshold values are related to easily identifiable turbulence characteristics. We focus our attention on the seeing that is certainly the most relevant parameter and for which it is easiest to find observations on other sites. In this section, we consider observations from VLT, LBT, TNG (La Palma), La Silla (Chile), Matera (Italy), Cagliari (Italy), and Kryoneri (Greece) for a total of seven sites (see Section \ref{obs} for characteristics on observations in the different sites). 

Figure \ref{RMSE_vs_mean_sigma} shows the RMSE$_{OBS-FOR}$ between observations and PC and observations and PP versus two statistical operators characterizing the seeing above the seven different sites: the median\footnote{We highlight that conclusions of this analysis are basically the same if we take the mean instead of the median.} and the standard deviation on a scale of 1h ($\sigma_{1h}$) of the seeing. We consider the standard deviation on a 1h timescale, as we study the forecast performance with FTS=1h. Therefore, it is more suitable to calculate the statistics of $\sigma$ on this timescale, instead of the whole duration of the dataset  (which has not provided interesting results).
Figure \ref{RMSE_vs_mean_sigma}-top shows the RMSE$_{OBS-FOR}$ between PC and PP versus the median of the seeing (left) and the $\sigma_{1h}$ (right) for the seven sites. Figure \ref{RMSE_vs_mean_sigma} at the bottom shows, for the PC case only\footnote{We find equivalent results for the PP case.}, the regression line (dashed black line) with respect to dots for the median case (left) and for the $\sigma_{1h}$ case (right).  Table \ref{table_correlation} reports for both cases PC and PP (FTS=1h) the Pearson coefficients and the RMSE between the dots and the regression lines. 

We observe that even if the RMSE$_{OBS-FOR}$ increases with respect to the median value as well as with $\sigma_{1h}$, in the case of $\sigma_{1h}$ we definitely have a higher Pearson coefficient and a lower RMSE value with respect to the median case. 
Looking at the bottom of Fig.\ref{RMSE_vs_mean_sigma} and Table \ref{table_correlation} we conclude therefore that $\sigma_{1h}$ is much better correlated to RMSE$_{OBS-FOR}$ than the median in the PC and the PP cases. This result is interesting, as it seems to indicate that it is not the typical seeing above a site that is better correlated to RMSE$_{OBS-FOR}$ between PC or PP and observations, but the intrinsic variability of the seeing on the same timescale on which RMSE$_{OBS-FOR}$ is calculated. 

We also observe that, as indicated by the blue shadow in Fig.\ref{RMSE_vs_mean_sigma} (bottom right), the dispersion of dots increases with the value of $\sigma_{1h}$. For $\sigma_{1h}$ < 0.15\arcsec, the correlation is much better, but for $\sigma_{1h}$ > 0.15\arcsec, we see a more evident dispersion.
It is worth noting that the error bars for Matera, Cagliari, and Kryoneri are smaller than those for the other sites, as we consider a smaller sample (one or two years instead of eight years). Certainly, the error bars should be larger for these three sites if we treated a sample of eight years, and considering that these are sites characterised by worse seeing, we should expect larger error bars than the other sites (particularly along the y-axis). Also, we observed that the post-processing procedure aiming to clean and filter the dataset for the three sites characterised by worse seeing is more delicate, as the treatment is more sensitive to filtering procedures. Finally, we repeated for the parameter $\sigma_{1h}$ the same calculation done for the RMSE (see Fig.\ref{resampling}) to ensure that there is no dependency on the sampling rate. We found that between a few seconds and around a minute the variation on $\sigma_{1h}$ is of the order of 0.01 \arcsec, which is definitely a negligible quantity in the context of Fig.\ref{RMSE_vs_mean_sigma} (bottom right).

The conclusion is that, at present, we cannot retrieve an analytical expression for the regression law, however,  we can state that a larger $\sigma_{1h}$ is certainly associated with a larger RMSE$_{OBS-FOR}$. That means that RMSE$_{OBS-FOR}$ are smaller for stable sites. We deduce, therefore, that it is much more difficult to provide useful forecasts for very stable sites as the RMSE$_{OBS-FOR}$ of the reference thresholds is much smaller. These results indicate that forecasting optical turbulence has significant potential in the field of free-space optical communication (FSOC), where site conditions are typically less favourable than at astronomical observatories. In FSOC cases we can easily be forced to apply the forecast in sites located close to cities and at low altitudes where the seeing is definitely worse than in the best astronomical sites.

\begin{table}
\caption{Correlation figures of merit.}
\centering
\begin{tabular}{l c c}\toprule
Figure of merit & PP & PC\\ \hline
median: Pearson &  0.717 & 0.692 \\
std: Pearson &  0.887 & 0.866 \\ 
median: RMSE & 0.080 & 0.084\\
std: RMSE & 0.053 & 0.058\\ \hline
\end{tabular}
\label{table_correlation}
\tablefoot{Values of different figures of merit quantifying the correlation between the RMSE$_{OBS-FOR}$ and the median and between RMSE$_{OBS-FOR}$ and $\sigma_{1h}$ on a rich statistical sample. These values are associated with Fig. \ref{RMSE_vs_mean_sigma}.}
\end{table}

\section{Conclusions}
\label{conc}
In this paper, we quantify the performance of two forecast methods (PC and PP) that represent references and offer a threshold when one performs forecasts of the average of a parameter in the coming  Y hours (PC) and when one performs the forecast of the temporal evolution of a parameter in the coming  Y hours. The value of  Y is the forecast timescale (FTS). We focused our attention on short FTS (1h and 2h), which are the most relevant for the science operations of ground-based facilities. Such an analysis is extremely relevant as it allows us to define a figure of merit that indicates whether the information provided by any forecasting method (numerical, statistical, or using neural networks or machine learning) is useful or not. In other words, if a forecasting method provides a performance worse than the associated PC or PP, it is useless. We find that these thresholds are site dependent; therefore, this means that any study providing OT forecast performance should be associated with a similar analysis to that in this paper to appreciate the impact of the forecasts provided.

In this paper we calculated these thresholds for the main astroclimatic parameters (seeing, isoplanatic angle, wavefront coherence time, and GLF) on a quite extended sample of measurements (8 years) above two of the best astronomical sites in the world where our group leads operational forecast systems of  OT to support the science operation of LBT and VLT (respectively, the projects ALTA and FATE). The threshold values for each parameter for FTS=1h and 2h are reported in Table \ref{table_results}. In Fig.\ref{seeing_PC}-Fig.\ref{Theta_PP}, the trends over an interval for FTS=[1h-4h] are shown.
In all cases, the threshold values for the PP and PC methods are different, with a variable difference depending on the parameter, on the site, on the FTS, and on the method to calculate the FTS. The analysis of atmospheric parameters is reported in Annex \ref{ann_d}.

Using observations extended on a richer number of sites, more precisely seven sites (see Section \ref{disc}), we also looked for the existence of any correlation of the RMSE$_{FOR-OBS}$ with typical characteristics of the OT of the sites themselves. This analysis aims to extrapolate the performance of forecast thresholds above third sites. We found that the median and mean values of the seeing are not a good figure of merit, while the median value of the standard deviation $\sigma$ calculated on a timescale of 1h $\sigma_{1h}$ (the same timescale for which forecasts are made) shows a much better correlation with RMSE$_{FOR-OBS}$ particularly for $\sigma_{1h}$ $\le$ 0.15\arcsec. Beyond this value, the dispersion of the data increases in such a way that it is not possible at this stage to provide an analytical expression for the correlation, but it is clear that a form of correlation exists between the two parameters. This result has an important consequence: the larger the  $\sigma_{1h}$, the higher the threshold for RMSE$_{FOR-OBS}$ seeing. In other words, it is easier for forecast models to bring added value when forecasting less stable sites. For example, it is easier to provide useful forecast information in free-space optical communication than in astronomical contexts, since the former are, in general, characterised by less stable atmospheric conditions than the latter. 

\begin{acknowledgements} 
We acknowledge the Large Binocular Telescope Observatory (LBTO), the European Southern Observatory (ESO), the Telescopio Nazionale Galileo (TNG), the Matera Space Centre, and the Miratlas company for access to their measurements. We thank Angel Otarola (ESO) and Steffen Mieske (ESO) for the support given for this study.
\end{acknowledgements} 

\bibliographystyle{aa}
\bibliography{references}

\begin{appendix}

\section{Observations: Statistics}
\label{ann_a}

\begin{figure}[h]
\includegraphics[width=\hsize]{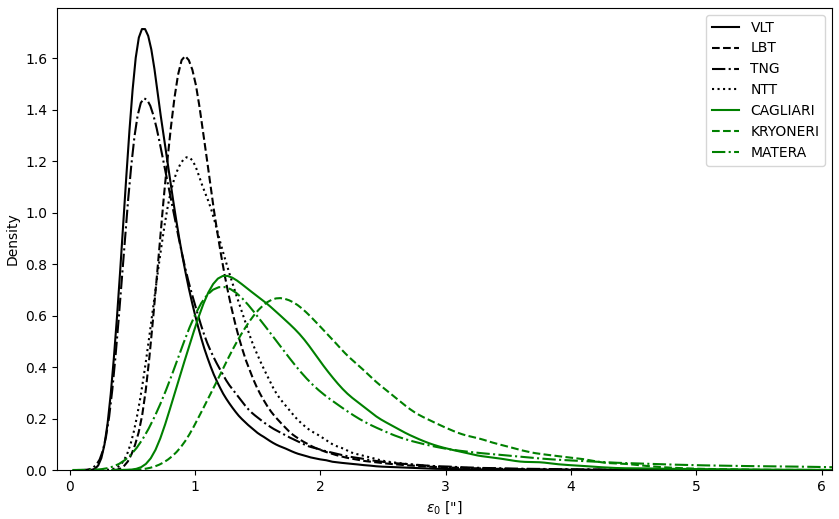}
  \caption{KDE (Kernel Density Estimation ) plot of seeing measures by density. Median values: 0.706" (VLT), 1.104" (LBT), 0.76" (TNG), 1.07" (La Silla), 1.5" (Matera), 1.57" (Cagliari), 1.91" (Kryoneri)}
     \label{KDEhist}
\end{figure}

We report in Fig. \ref{KDEhist} and Table \ref{table_stats} the characterisation of the seeing on the different sites analysed in the respective data-set. For VLT, LBT and TNG we have the [2017-2024] range (8 years); for La Silla site we have the [2017-2023] range (7 years) for Matera we have the [2022/01/01-2022/12/31] range (1 year), for Cagliari and Kryoneri we have the [2024/01/01,2025/12/31] range (2 years). Fig. \ref{KDEhist} shows the Kernel Density Estimation (KDE) plot for all the sites analysed in this paper. 

\begin{table}[h]
\caption{Site measurement statistics}
\label{table_stats}
\centering                       
\begin{tabular}{c c c c c}
\hline\hline                 
Parameter & mean & median & $\sigma$ & skew \\
\hline                        
   \bf{VLT} \\
   $\varepsilon$ [``] & 0.81 & 0.71 & 0.39 & 2.07\\
   $\tau_0$ [ms] & 5.38 & 4.59 & 3.26 & 1.72\\
   $\theta_0$ [``] & 2.06 & 1.98 & 0.74 & 0.91\\
   GLF [\%] & 61.9 & 64.7 & 19.1 & -67.9\\
   $PWV_{night}$ [mm] & 3.30 & 2.44 & 2.69 & 1.98\\
   $PWV_{day}$ [mm] & 3.69 & 2.73 & 2.94 & 1.76\\
   $WS_{night} [ms^{-1}]$& 6.47 & 5.88 & 3.86 & 0.64\\
   $WS_{day} [ms^{-1}]$& 7.08 & 6.18 & 4.25 & 1.00\\
   $T_{night}$ [°C]& 12.97 & 13.3 & 2.76 & -0.58\\
   $T_{day}$ [°C]& 13.58 & 13.86 & 2.81 & -0.70\\
   $RH_{night}$ [\%]& 14.3 & 10.0 & 11.8 & 1.77\\
   $RH_{day}$ [\%]& 17.4 & 13.0 & 13.8 & 1.70\\
   \bf{LBT} \\
   $\varepsilon$ [``] & 1.10 & 1.01 & 0.40 & 2.27\\
   $WS_{night} [ms^{-1}]$& 7.31 & 6.44 & 4.39 & 1.08\\
   $WS_{day} [ms^{-1}]$& 6.99 & 5.96 & 4.76 & 1.07 \\
   $T_{night}$ [°C]& 3.93 & 3.9 & 6.61 & -0.21\\
   $T_{day}$ [°C]& 5.59 & 6.0 & 7.23 & -0.27\\
   $RH_{night}$ [\%]& 45.3 & 41.4 & 26.9 & 0.43\\
   $RH_{day}$ [\%]& 45.9 & 40.2 & 26.8 & 0.52\\
   \bf{TNG} \\
   $\varepsilon$ [``] & 0.91 & 0.77 & 0.52 & 2.21\\
   \bf{NTT} \\
   $\varepsilon$ [``] & 1.17 & 1.07 & 0.44 & 1.62 \\
   \bf{Matera}\\
   $\varepsilon$ [``]& 1.86 & 1.5 & 1.20 & 2.52\\
   \bf{Cagliari} \\
   $\varepsilon$ [``] & 1.71 & 1.57 & 0.69 & 1.48 \\
   \bf{Kryoneri} \\
   $\varepsilon$ [``] & 2.07 & 1.91 & 0.75 & 1.08 \\
   
\hline                                   
\end{tabular}
\tablefoot{The measurements span years 2017-2024  for VLT, LBT, and TNG, 2017-2023 for La Silla, 2022 for Matera, 2024-2024 for Cagliari and Kryoneri}
\end{table}

\section{Parameters definition}
\label{ann_b}
We report here the analytical expressions of the astroclimatic parameters studied in this paper. In Eq.\ref{eq1} the Fried parameter $r_{0}$ measured in centimetres and normalized with respect to the zenith:

\begin{equation}
r_{0}=\left( 0.423\cdot\left( \frac{2\pi}{\lambda} \right)^{2} \int_{0}^{ \infty }\mathrm{C}_{N}^{2}(h)dh)\right)^{-3/5}
\label{eq1}
\end{equation}

where $\lambda$ is the wavelength and h is the height above the ground. From Eq.\ref{eq1} we can retrieve the seeing ($\epsilon$) that is the width at the half height of the Point Spread Function (PSF) at the focus of the telescope with D > $r_0$. The analytical expression normalised with respect to the zenith and measured in arcseconds is reported in Eq.\ref{eq2}:

\begin{equation}
\varepsilon=0.98\frac{\lambda}{r_0}
\label{eq2}
\end{equation}

The isoplanatic angle ($\theta_0$) is the angular equivalent of the Fried parameter, it is the maximum angular separation between two stellar objects such that the mean-squared wavefront error is 1 rad$^2$. The analytical expression normalized with respect to the zenith is given by Eq.\ref{eq3} and measured in arcseconds:

\begin{equation}
\theta_{0}=0.057\cdot\lambda^{6/5}\left(\int_{0}^{\infty }h^{5/3}\mathrm{}C_{N}^{2}(h))  \right)^{-3/5}
\label{eq3}
\end{equation}

The wavefront coherence time ($\tau_{AO}$) tells us how fast the turbulence is. The analytical expression normalised with respect to the zenith is given by Eq.\ref{eq4} and measured in milliseconds:

\begin{equation}
\tau_{0}=0.057\cdot\lambda^{6/5}\left( \int_{0}^{\infty} |\mathbf{V}(h)|^{5/3} C_N^2(h)dh \right)^{-3/5}
\label{eq4}
\end{equation}

where V(h) is the horizontal wind velocity at the height h. In a multilayer atmosphere $\tau_{0}$ is the time required to pass across r$_{0}$ for  a turbulence layer equivalent to the whole 20~km and having the velocity equal to V$_{0}$ (Eq.\ref{eq5}):

\begin{equation}
V_{0}=\left[ \frac{\int_{0}^{\infty }\left|V(h))  \right|^{5/3}\mathrm{C}_{N}^{2}dh}{\int_{0}^{\infty}\mathrm{C}_{N}^{2}dh} \right]^{3/5}
\label{eq5}
\end{equation}

The mixing ratio of water vapour is $MR = \frac{m_{wp}}{m_d}$ [g/kg] where $m_{wp}$ is the mass of water vapour and $m_{d}$ is the corresponding mass of dry air. The relative humidity is given by $RH = 100*\frac{MR}{MR_s}$, where $MR_s$ is defined as the mass of water vapour in a given volume of saturated air $m_{vs}$ with respect to the mass of dry air $m_d$.
\\
The precipitable water vapour (PWV expressed in millimetres) is the total atmospheric water vapour contained in a vertical column of unit cross-sectional area extending between any two specified levels (e.g., from the earth's surface to the top of the atmosphere). It is commonly expressed in terms of the height to which that water would stand if completely condensed and collected in a vessel of the same unit cross section. Expression of PWV is given by Eq.\ref{eq6}:

\begin{equation}
PWV = \frac{10^3}{\rho g}\int^{p_{sup}}_{p_{inf}}MR_{dp}
\label{eq6}
\end{equation}
where MR is the mixing ratio of the water vapour, p is the atmospheric pressure in, p$_{inf}$ is the pressure at 20m, p$_{sup}$ is the pressure at 20km, $\rho$ is the water density, g is the gravity constant in (ms$^{-2}$).

\section{Analysis for atmospheric parameters}
\label{ann_d}

We report here the results of the analysis performed in Section \ref{ana}, but for the atmospheric parameters (precipitable water vapour, temperature, relative humidity and wind speed) above VLT and LBT for the PC and the PP. 

\begin{figure*}[h]
  \centering
  \includegraphics[width=0.87\hsize]{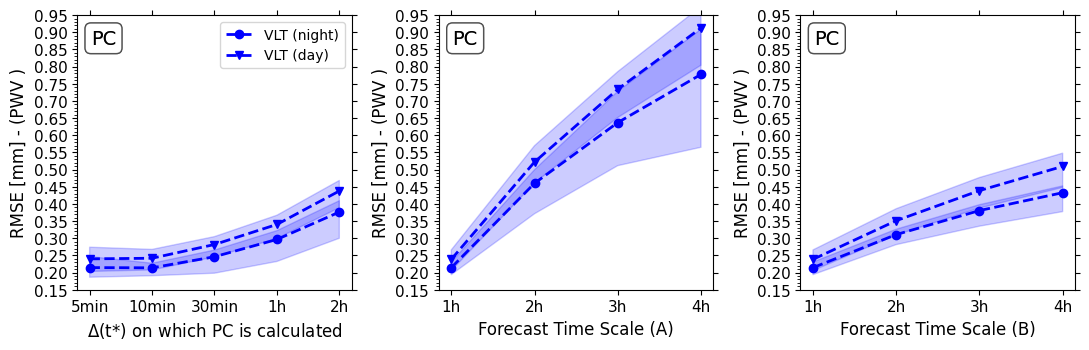}
  \caption{Same as Fig.\protect\ref{seeing_PC} but for the precipitable water vapour.  The optimal  $\Delta(t^*)$  is 5 minutes.}
  \label{PWV_PC}
\end{figure*}

\begin{figure*}
  \centering
  \includegraphics[width=0.87\hsize]{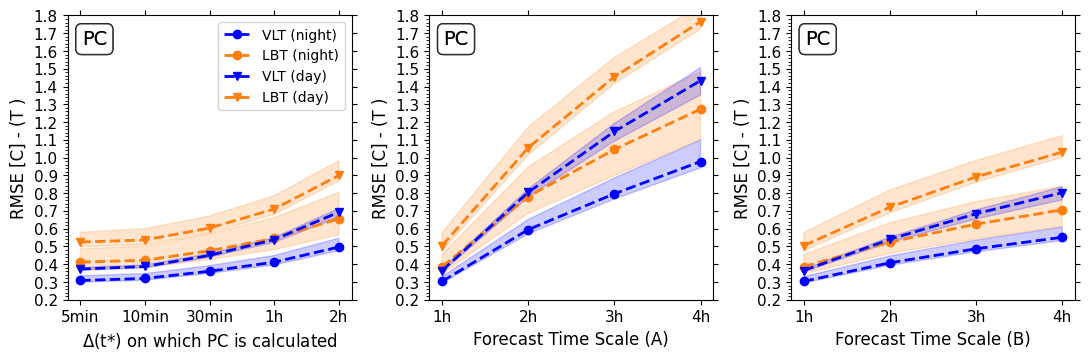}
  \caption{Same as Fig.\protect\ref{seeing_PC} but for the temperature.  The optimal  $\Delta(t^*)$  is 5 minutes.} 
  \label{T_PC}
\end{figure*}

\begin{figure*}
  \centering
  \includegraphics[width=0.87\hsize]{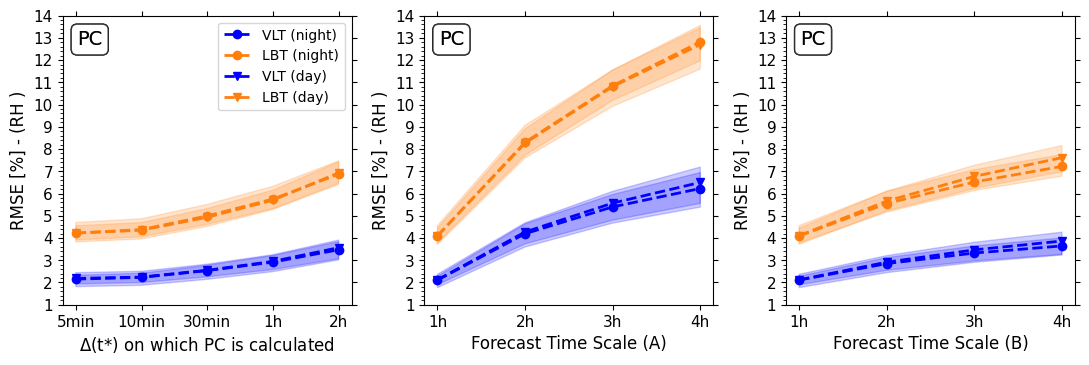}
  \caption{Same as Fig.\protect\ref{seeing_PC} but for the relative humidity.  The optimal  $\Delta(t^*)$  is 5 minutes.}
  \label{RH_PC}
\end{figure*}

\begin{figure*}
  \centering
  \includegraphics[width=0.87\hsize]{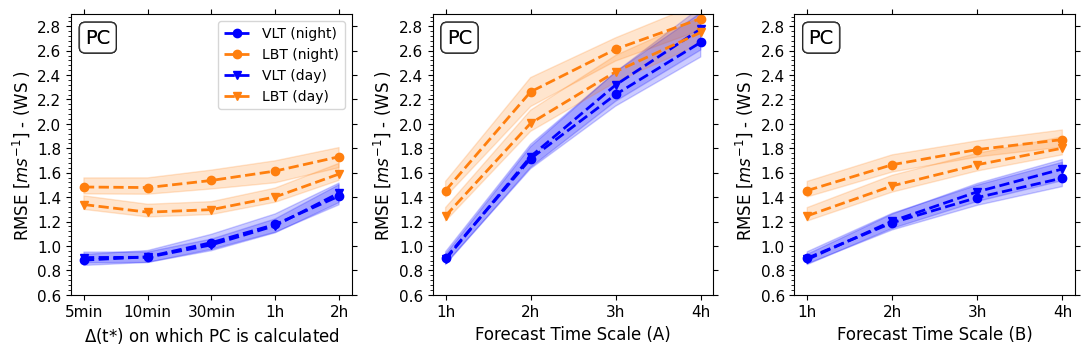}
  \caption{Same as Fig.\protect\ref{seeing_PC} but for the wind speed.  The optimal  $\Delta(t^*)$  is 10 minutes.}
  \label{WS_PC}
\end{figure*}

\begin{figure*}
  \centering
  \includegraphics[width=0.87\hsize]{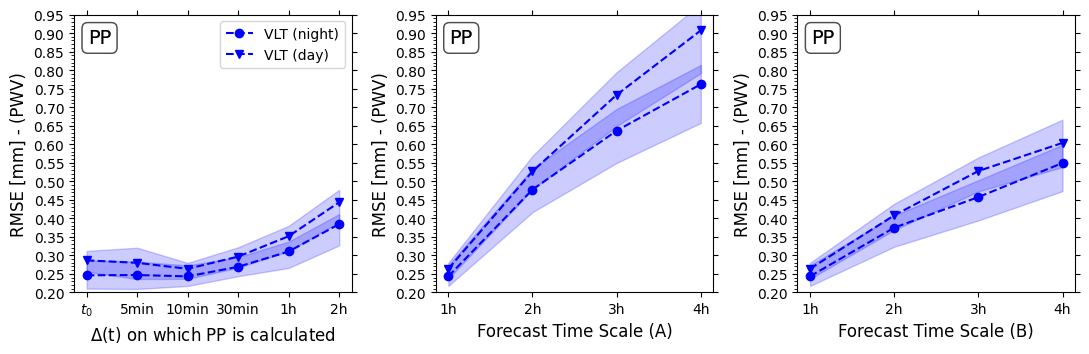}
  \caption{Same as Fig.\protect\ref{seeing_PP} but for the precipitable water vapour.  The optimal  $\Delta(t)$  is 10 minutes}
  \label{PWV_PP}
\end{figure*}

\begin{figure*}
  \centering
  \includegraphics[width=0.87\hsize]{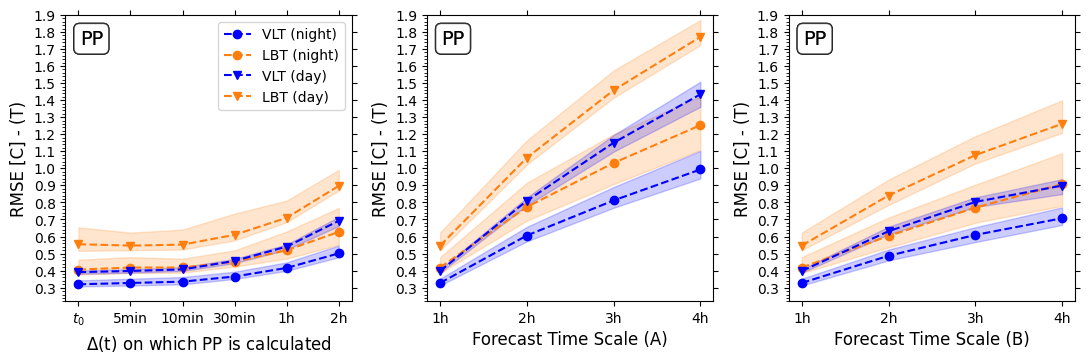}
  \caption{Same as Fig.\protect\ref{seeing_PP} but for the temperature.  The optimal  $\Delta(t)$  is 10 minutes}
  \label{T_PP}
\end{figure*}

\begin{figure*}
  \centering
  \includegraphics[width=0.86\hsize]{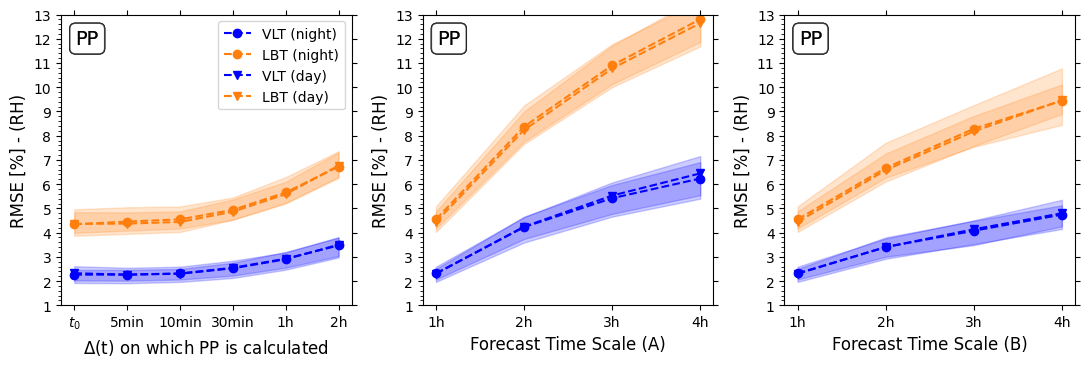}
  \caption{Same as Fig.\protect\ref{seeing_PP} but for the relative humidity.  The optimal  $\Delta(t)$  is 10 minutes}
  \label{RH_PP}
\end{figure*}

\begin{figure*}
  \centering
  \includegraphics[width=0.86\hsize]{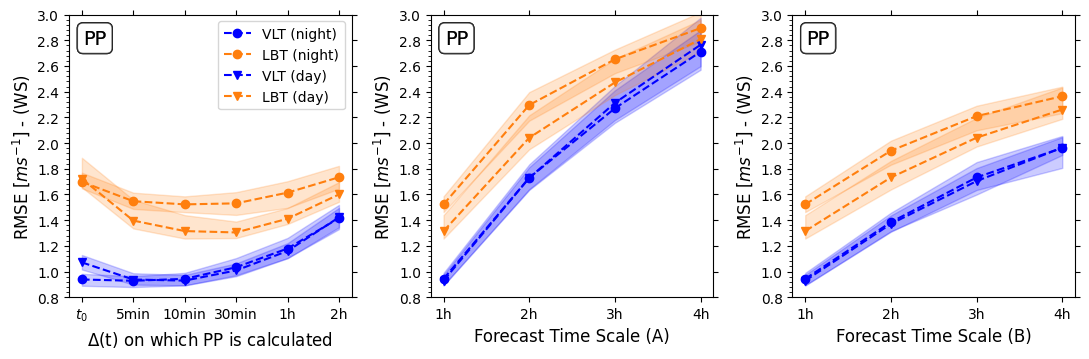}
  \caption{Same as Fig.\protect\ref{seeing_PP} but for the wind speed.  The optimal  $\Delta(t)$  is 10 minutes}
  \label{WS_PP}
\end{figure*}

Looking at Fig.\ref{PWV_PC}-Fig.\ref{WS_PP}, left column we conclude that the optimal $\Delta$t = 10~min for basically all the parameters. We observe that, in the same way as for the astroclimatic parameters, the RMSE$_{FOR-OBS}$ is smaller above VLT than above LBT for almost all the timescales. Only for the WS for FTS=4h we have a similar RMSE above the two sites. For the atmospherical parameters we distinguish between night-time and daytime. Interestingly, for the T we have a larger RMSE$_{FOR-OBS}$ at a FTS=1h for the daytime with respect to the nighttime while for the WS we have the opposite trend in both cases (PP and PC) at least above LBT. 

The sensors at the LBT are located on top of the dome (height around 40~m).  As reported in \citep{turchi2017} the anemometers for the WS are at 3~m (front) and 5~m (back) above the dome. Sensors for temperature T and relative humidity RH are also located in the back mast.
The sensors at the VLT are located at  30~m above the ground level (a.g.l.) (see ESO archive repository). That means that at VLT sensors are located closer to the ground than LBT and the $\Delta$H between the VLT and LBT is of the order 10-15 meters. This might also explain the slightly larger RMSE$_{FOR-OBS}$ at LBT relative to VLT for T, RH, and WS. 

\clearpage

\end{appendix}
\end{document}